\documentclass[%
reprint,
superscriptaddress,
pre,
]{revtex4-1}
\usepackage{ amssymb,amsmath}

\usepackage{graphicx}

\providecommand{\acknowledgments}{\section*{Acknowledgments}}

\usepackage[
colorlinks=true,
linkcolor=blue,
citecolor=blue,
filecolor=blue,
urlcolor=blue]{hyperref}

\usepackage{sidecap}
\sidecaptionvpos{figure}{t}

\begin{document}

\title{Nonlinear dispersive waves in repulsive lattices}

\author{A.~Mehrem}
\affiliation{Instituto de Investigaci\'{o}n para la Gesti\'{o}n,
Integrada de las Zonas Costeras, Universitat Politecnica de
Valencia, Paranimf 1,  46730 Grao de Gandia, Spain}

\author{N.~Jim\'{e}nez}
\affiliation{LUNAM Universit\'e, Universit\'e du Maine, CNRS, LAUM UMR 6613, Av. O. Messiaen, 72085 Le Mans, France}

\author{L.~J.~Salmer\'{o}n-Contreras}
\affiliation{Instituto Universitario de Matem\'{a}tica Pura y
Aplicada, Universitat Politecnica de Valencia, Camino de Vera
s/n, 46022 Valencia, Spain}

\author{X.~Garc\'{i}a-Andr\'{e}s}
\affiliation{Departamento de Ingenieria Mecanica y Materiales, Universitat Politecnica de Valencia, Camino de Vera s/n, 46022 Valencia, Spain}

\author{L.~M.~Garc\'{i}a-Raffi}
\affiliation{Instituto Universitario de Matem\'{a}tica Pura y
Aplicada, Universitat Politecnica de Valencia, Camino de Vera
s/n, 46022 Valencia, Spain}%

\author{R.~Pic\'{o}}
\affiliation{Instituto de Investigaci\'{o}n para la Gesti\'{o}n,
Integrada de las Zonas Costeras, Universitat Politecnica de
Valencia, Paranimf 1,  46730 Grao de Gandia, Spain}

\author{V.~J.~S\'{a}nchez-Morcillo}
\affiliation{Instituto de Investigaci\'{o}n para la Gesti\'{o}n,
Integrada de las Zonas Costeras, Universitat Politecnica de
Valencia, Paranimf 1,  46730 Grao de Gandia, Spain}

\date{\today}

\begin{abstract}
The propagation of nonlinear waves in a lattice of repelling particles is studied theoretically and experimentally. A simple experimental setup is proposed, consisting of an array of coupled magnetic dipoles. By driving harmonically the lattice at one boundary, we excite propagating waves and demonstrate different regimes of mode conversion into higher harmonics, strongly influenced by dispersion and discreteness. The phenomenon of acoustic dilatation of the chain is also predicted and discussed. The results are compared with the theoretical predictions of $\alpha$-FPU equation, describing a chain of masses connected by nonlinear quadratic springs and numerical simulations. The results can be extrapolated to other systems described by this equation.  
\end{abstract}

\maketitle

\section{Introduction }

Repulsive interactions among particles are known to form ordered states of matter. One example is Coulomb interaction, which is on the basis of the solid state physics \cite{Kittel}. In a crystal, atoms and ions are organized in ordered lattices by means of repulsive forces acting among them. Such non-contact forces provide also the coupling between neighbour atoms, what allows the propagation of perturbations in the form of phonons, or elementary excitations of the lattice. This picture is not restricted to the atomic scale. At a higher scale, the interaction of charged particles other from atoms and ions has shown the formation of crystal lattices. A remarkable case is ionic crystals in a trap \cite{raizen1992ionic}. Such crystals, which are considered a particular form of condensed matter, are formed by charged particles, e.g. atomic ions, confined by external electromagnetic potentials (Paul or other traps), and interacting by means of the Coulomb repulsion. Crystallization requires low temperature that is achieved by laser cooling techniques. Different crystallization patterns have been observed by tuning the shape and strengths of the traps. Crystals of trapped ions have been subject of great attention as a possible configuration to perform quantum computation \cite{porras2008mesoscopic}. Crystallization of a gas of confined electrons, known as Wigner crystals, have been also predicted and observed \cite{Matveev10,Deshpande08}. 

Waves in such crystals show strong dispersion at wavelengths comparable with the lattice periodicity. The linear (infinitesimal amplitude) dispersion relation, and some nonlinear characteristics of wave propagation have been experimentally determined in an electrically charged, micrometer-sized dust particles immersed in the sheath of a parallel plate rf discharge in helium in a rf plasma
\cite{homann1997determination}, where the waves are excited by transferring the momentum from a laser to the first particle in the chain.

In other type of plasma crystals, linear wave mixing and harmonic generation of compressional waves has been theoretically \cite{avinash2003nonlinear}
and experimentally \cite{nosenko2004nonlinear} demonstrated. Also, nonlinear standing waves have been discussed in a two-dimensional system of charged particles \cite{denardo1988theory}. Here the generation of second and third harmonics was predicted on the long-wavelength (non- or weakly dispersive) limit.

Some experiments with analogue models of repulsive lattices have been done using magnets as interacting particles, with the aim of demonstrating the generation and propagation of localized perturbations (discrete breathers and solitons). For example, in the seminal work of Russel
\cite{russell1997moving} a chain of magnetic pendulums (very similar to the setup presented in this paper) was used to simulate at the macroscopic level some natural layered silicate crystals, such as muscovite mica. More recently, in \cite{moleron2014solitary}, the authors proposed another configuration of a chain of repelling magnets, for the study of solitary waves, similar to the highly discrete kinks studied theoretically in Coulomb chains including realistic interatomic and substrate potentials \cite{archilla2015ultradiscrete}.

We finally note that repulsive potentials are not restricted to those of electric or magnetic nature. A celebrated case is the granular chain of spherical particles interacting via Hertz potentials. Many studies have been done in this system, theoretical and experimental, on the propagation of the solitary waves. Recently, several nonlinear effects related to the propagation of intense harmonic waves in such granular lattices has been described in \cite{sanchez2013second}, with special attention to the dispersive regime. 

In this work, we investigate experimentally and numerically the propagation of nonlinear and dispersive waves in harmonically driven repulsive lattices with on-site potentials. In particular, we study the harmonic generation of monochromatic waves travelling in an array of coupled magnetic dipoles, comparing the observations with the predictions from the $\alpha$-FPU equation and numerical results including an on-site potential. Two main results are reported: first, the experimental observation of the generation of second harmonic in highly dispersive nonlinear lattices and, second, the saturation in the generation of the evanescent zero frequency mode in lattices with on-site potential. The paper is organized as follows: In Sec.~\ref{sec:theory}, the theoretical model, the equation of motion of a lattice of particles interacting by inverse power-law forces, is presented. The weakly nonlinear limit is considered, where the model approaches to the celebrated $\alpha$-FPU equation. The linear dispersion relation and the analytical solutions for propagating and evanescent nonlinear periodic waves, are given. In Sec.~\ref{sec:magnets}, the theory is particularized for the case of an array of coupled magnetic pendula, and it is presented the experimental setup based on a lattice of magnetic pendula rotating by means of a magnetic bearing system that guarantees low friction. In Sec.~\ref{sec:results} we discuss the experimental results, concerning the generation of harmonics and a static displacement (dilatation) mode. Finally, the conclusions of the study are given in Sec.~\ref{sec:conclusions}.

\section{Theoretical model}\label{sec:theory}
\subsection{Equation of motion}
\begin{figure}[b]
	\begin{center}
		\includegraphics[width=0.90\columnwidth]{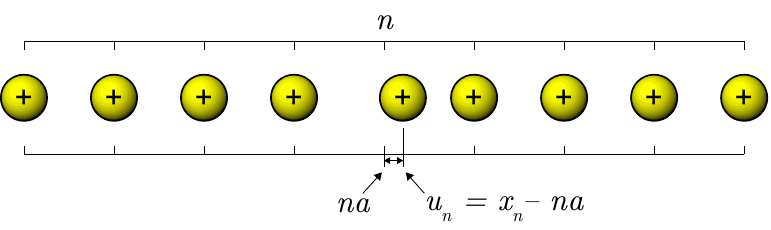}
		\caption{Scheme of the lattice of non-linearly coupled oscillators.}
		\label{schem2}
	\end{center}
\end{figure}
We consider an infinite chain of identical particles with mass $M$ aligned along the $x$-axis, interacting with their nearest neighbours via repulsive potentials, $V_\mathrm{int}$. In the absence of perturbations, every mass has a fixed equilibrium position, with the interparticle distance given by $a$, as shown in Figure $\ref{schem2}$. Since the forces are repulsive, note that for a finite chain this is only possible if there is an external potential $V_\mathrm{ext}$ that keeps the particles confined. This effect can be provided by a periodic on-site potential, or a force keeping the boundary particles at fixed positions. The equation of motion can be written as
\begin{equation}
M \ddot{u}_n=V'_\mathrm{int}\left(u_{n+1}-u_{n}\right)
            -V'_\mathrm{int}\left(u_{n}-u_{n-1}\right) +V'_\mathrm{ext},
            \label{eq:eqmotion}
\end{equation}
\noindent where $u_n$ stands for the displacement of the $n$-th particle measured with respect to its equilibrium position, $M$ is the mass of the particle, $V$ are the potentials and $V'$ their derivatives with respect to the spatial coordinate, i.e., the forces. For small displacements, the interaction forces, $V'_\mathrm{int}$, can be considered linear with respect to the distance between the particles, $r$, i.e., $V'(r)=\kappa r$ where $\kappa$ is a constant, then Eq. (\ref{eq:eqmotion}) represents a system of coupled harmonic oscillators. For higher amplitude displacements, the linear approximation of the interaction force cannot be assumed in most real systems and nonlinearity must be considered. Chains of nonlinearly coupled oscillators have been extensively studied in the past for different types of anharmonic interaction potentials. Some relevant cases are the $\alpha$-FPU lattice, where $V'(r)=\kappa_1 r+\kappa_2 r^2$ (quadratic interaction), the $\beta$-FPU lattice where $V'(r)=\kappa_1 r+\kappa_3 r^3$ (cubic interaction), the Toda lattice, with $V'(r)=\exp({-r})-1$ or the granular lattice, with $V'(r)=\kappa r^{3/2}$. Here we consider the case of forces that decrease with an inverse power law of the distance, $V'(r)=\beta r^{-\alpha}$, typical of interatomic interactions, e.g. as the Coulomb repulsive interaction. For such a force, the equation of motion results in
\begin{equation} \label{eq:eqmotion2}
	M \ddot{u}_n= \frac{\beta}{(a-u_{n+1}+u_{n})^\alpha}-\frac{\beta}{(a-u_{n}+u_{n-1})^\alpha} +V'_\mathrm{ext}.
\end{equation}

The exponent $\alpha$ can take different values depending on the particular system: $\alpha=2$ for electrically charged particles, e.g. in ion Coulomb crystals \cite{raizen1992ionic} or dusty plasma crystals \cite{nosenko2004nonlinear}, $\alpha=4$ for distant magnetic dipoles \cite{russell1997moving}, or any other non-integer power \cite{moleron2014solitary}.  

In general, Eq.~(\ref{eq:eqmotion2}) do not possess analytical solutions. Approximate analytical solutions can be obtained in the small amplitude limit, i.e., assuming that the particle displacement $|u_n|$ is small compared to the lattice constant $a$. Under this assumption, the forces can be expanded in Taylor series and Eq.(\ref{eq:eqmotion2}) can be reduced, neglecting cubic and higher order terms, to an equation in the normalized form
\begin{align}\label{eq:FPU}
\ddot{u}_n=&\frac{1}{4}\left(u_{n-1}-2u_{n}+u_{n+1}\right)- \nonumber \\ 
           &\frac{\varepsilon}{8}\left(u_{n-1}-2u_{n}+u_{n+1}\right)\left(u_{n-1}-u_{n+1}\right) + \Omega_0^2 u_n,
\end{align}
\noindent where the normalization $u_n=u_n/a$ has been introduced, dots indicate now derivative with respect to a dimensionless time $\tau=\omega_m t$, where $\omega_m=\sqrt{4 \alpha \beta/M a^{\alpha+1}}$ is the maximum frequency of propagating waves (upper cutoff frequency of the dispersion relation), $\varepsilon=(1+\alpha)u_0$ is the nonlinearity coefficient and $\Omega_0=\omega_0/\omega_m$ is the on-site potential characteristic frequency. The on-site restoring force $V'_\mathrm{ext}$ is in general nonlinear. However, for small displacements, as considered here, it may be represented by a term $V'_\mathrm{ext}=M \Omega_0^2 {u_n}{a}$, where $\Omega_0$ is related with the frequency of oscillation of the particle in the external potential. The particular form of this term for the proposed experimental setup will be discussed later. If the on-site potential term is neglected (no external forces acting on the chain), Eq.~(\ref{eq:FPU}) reduces to the celebrated $\alpha$-FPU equation. It has been considered as an approximate description of many different physical systems, and has played a central role in the study of solitons and chaos \cite{GavallottiFPU}. 

\subsection{Dispersion Relation}
Some important features of the propagation of waves in a lattice can be understood by analyzing its dispersion relation. For infinitesimal amplitude waves, it can be obtained analytically by neglecting the nonlinear terms in the equation of motion, and solving for a harmonic discrete solution in the form $u_n = \exp {i(\Omega t-k n)}$, where $\Omega=\omega/\omega_m$ is the normalized wave frequency and $k$ is the wavenumber. By replacing this solution in the linearized Eq.~(\ref{eq:FPU}), we obtain the well-known dispersion relation for a monoatomic lattice, that in normalized form reads 
\begin{equation}\label{disp2}
	\Omega= \sqrt{\sin^2\left(\frac{k}{2}\right)+\Omega_0^2}.
\end{equation}
\noindent On one hand, there is an upper cutoff frequency at which the transition from propagative to evanescent solutions is produced, i.e., $\mathrm{|Im(k)|>0}$, and it is given in this normalization by $\Omega=\sqrt{1+\Omega_0^2}$. On the other hand, the effect of the on-site potential is to create a low frequency bandgap in the dispersion relation, i. e., $\Omega_0$ represents the lower cut-off frequency. In the absence of external confining potential, $\Omega_0 \rightarrow 0$, the dispersion relation reduces to $\Omega= \left| \sin\left({ka}/{2}\right) \right|$. In this case the upper cutoff normalized frequency is $\Omega=1$. 

Although the dispersion relation has been derived assuming infinitesimal amplitude (linear) waves, it describes also the propagation of other modes as the higher harmonics of a fundamental harmonic wave (FW), when these are generated by weakly nonlinear processes, as described in the following sections. 

\subsection{Analytical solutions}
One known effect of the quadratic nonlinearity is the generation of second and higher harmonics of an input signal. This is the basic effect, for example, of nonlinear acoustic waves propagating in  homogeneous, non-dispersive media \cite{naugolnykh1998nonlinear,hamilton1998nonlinear}, where the amplitude of the harmonics depends on the nonlinearity of the medium, the excitation signal and the propagated distance (the excitation amplitude in the chain $u_0$, the frequency $\Omega$ and its position $n$). 

In general, the generation of harmonics is strongly dependent of the dispersion of the system, as occurs in the discrete lattice described by Eq.~(\ref{eq:FPU}). To study the process of harmonic generation, an analytical solution can be obtained by perturbative techniques, such as the successive approximations method. We follow this approach, by assuming that the nonlinear parameter $\varepsilon$ is small (which implies displacements much smaller than the interparticle separation), and expressing the displacement as a power series in terms of $\varepsilon$, in the form $u_n=u_n^{(0)}+\varepsilon u_n^{(1)}+\varepsilon^2 u_n^{(2)}+\ldots$. After substituting the expansion into Eq.~(\ref{eq:FPU}), and collecting terms at each order in $\varepsilon$, we obtain a hierarchy of linear equations that can be recursively solved. This has been done in Ref.~\cite{sanchez2013second} to study nonlinear waves in a granular chain, formed by spherical particles in contact interacting by Hertz potentials, and the result is readily extendible to chain of particles interacting by inverse power laws of arbitrary exponent, which results in a particular value of the nonlinearity coefficient. The equation of motion is always given by Eq.~(\ref{eq:FPU}), the value of $\varepsilon$ being dependent on the exponent $\alpha$.  
In case of granular chain, it was shown that $\varepsilon=u_0/2$. In this work a chain with quasi-dipolar interaction, $\alpha=4$, gives $\varepsilon=5u_0$, i.e. the nonlinear effects are one order of magnitude higher. 

\begin{figure*}
	\centering
	\includegraphics[width=0.99\textwidth]{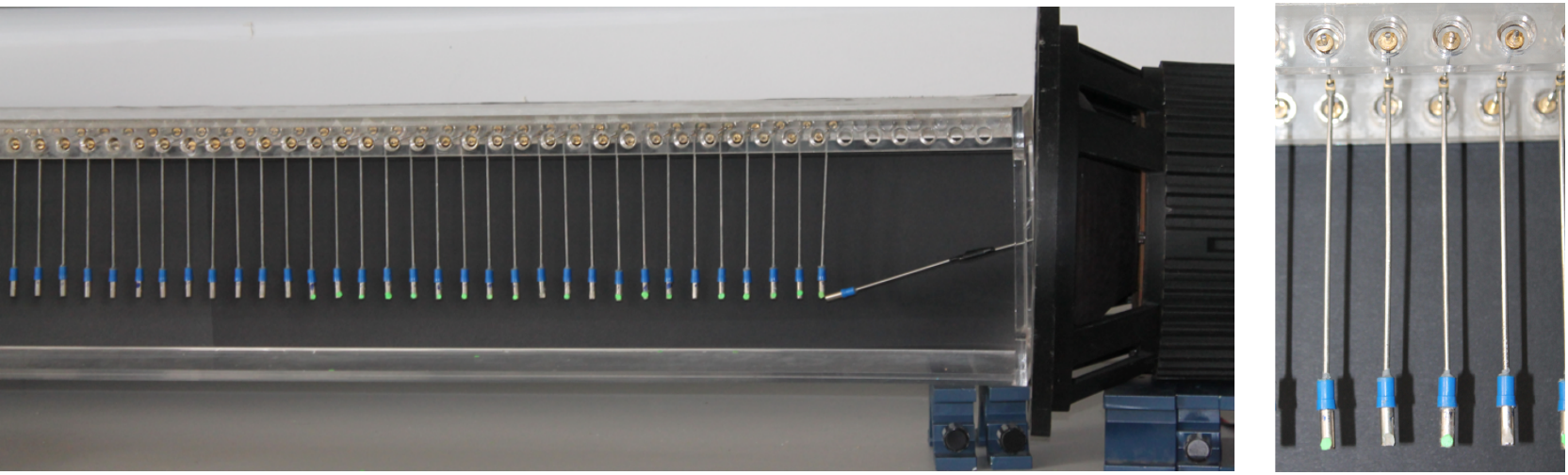}
	\caption{Photograph of the experimental setup. The chain of magnets is driven mechanically by a dynamic sub-woofer speaker. On right, a detail of the construction of the pendula with the magnetic quasi-levitation system for minimize the losses at the bearing.}
	\label{setup}
\end{figure*}

Up to second order of accuracy in $\varepsilon$, the displacement field can be expressed as (the details of the derivation can be found in Ref. \cite{sanchez2013second}) 
\begin{align} \label{eq:Analytic}
	u_n=&\varepsilon \Omega^2 n + \nonumber\\ 
		&\frac{1}{2}\left[1+\frac{1}{4} i  \varepsilon^2 C_\Omega \sin\left(\frac{\Delta k}{2}n\right) e^{i\frac{\Delta k}{2}n}\right] e^{i\theta_n}+\nonumber\\
		&\frac{\varepsilon}{4} \cot\left(\frac{k}{2}\right) \sin\left(\frac{\Delta k}{2}n\right) e^{i\frac{\Delta k}{2}n} e^{2i\theta_n} + \mathrm{c.c.},
\end{align}
\noindent where $\theta_n=\Omega t-kna$, $n$ is the oscillator number corresponding to the discrete propagation coordinate, and
\begin{equation}\label{eq:Di1}
	C_\Omega=1-\frac{\sin[k(2\Omega)/2]}{ \sin[\Delta k(\Omega)/2]},
\end{equation}
\noindent where $\Delta k=2 k(\Omega)-k(2 \Omega)$ is the wavenumber mismatch between the forced, $2k(\Omega)$, and free, $k(2\Omega)$, contributions to the second harmonic. 

The solution given by Eq.~(\ref{eq:Analytic}) describes wave propagation in the system when the frequency of the second harmonic belongs to the dispersion relation, which is the case for driving frequencies $\Omega< 1/2$. For higher driving frequencies, the second harmonic frequency is outside the propagation band (becoming an evanescent mode) and the solution takes the form
\begin{align} \label{eq:Analytic2}
	u_n = &\varepsilon \Omega^2 n + \nonumber \\
	&\frac{1}{2}\left[1+\frac{1}{8} \varepsilon^2 C_\Omega \left(1-e^{-k{''}n}e^{i k{'}n}\right)\right] e^{i\theta_n}+ \nonumber \\
	&\frac{\varepsilon}{8} \cot \left(\frac{k}{2}\right) \left(1-e^{-k{''}n}e^{i k{'}n}\right) e^{2i\theta_n} + \mathrm{c.c.},
\end{align}
\noindent where $k{''}=2 \cosh^{-1}(2\Omega)$ and $k{'}=2 k(\Omega)-1$ and the mismatch take now the form $\Delta k=k{'}+i k{''}$.

The previous analytical solutions (\ref{eq:Analytic}-\ref{eq:Analytic2}) predict a number of distinctive features in the nonlinear dynamics of the system, depending on the frequency regime. In the case of the second harmonic belonging to the propagation band, Eq.~(\ref{eq:Analytic}), dispersion causes a beating in the amplitudes of the different harmonics, since two components of the second harmonic with different wavenumbers propagate asynchronously. Both, the fundamental wave and its second harmonic oscillate out of phase in space: the displacement of the fundamental is maximum where the second harmonic vanishes, which occurs at positions satisfying the following condition: $n=2 \pi/\Delta k$. This process repeats periodically in space as energy is transferred between the two waves as they propagate. The half distance of the spatial beating period corresponds to the coherence length $l_c$: 
\begin{equation}\label{coherence}
l_c =\frac{\pi}{\Delta k},
\end{equation}   
\noindent  and it physically corresponds to the position where the free and forced waves are exactly in phase, i.e., the location of the maximum of first spatial beat.

When the second harmonic frequency lies beyond the cut-off frequency, the free wave is evanescent. There still exists however a forced wave, driven by the first harmonic at any point in the chain. Due to this continuous forcing, the amplitudes of the fundamental and its second harmonic do not oscillate, reaching the amplitude of the second harmonic a constant value after a short transient of growth. This implies propagation of the second harmonic even in the forbidden region.
We note that similar results about the behaviour of harmonics have been obtained for nonlinear acoustic waves propagating in a 1D periodic medium or superlattice \cite{jimenez2016nonlinear}.
Finally, we note that the theory predicts the existence of a zero-frequency mode, $u_n=\varepsilon \Omega^2 n$, which represents an static deformation of the lattice, i.e., a constant dilatation. This effect will be studied in detail in Sect. IV.

\section{The lattice of magnetic dipoles}\label{sec:magnets}
\subsection{Forces acting on a magnet}
Consider two magnetic dipoles, with magnetic moments $\vec{m}_1$ and $\vec{m}_2$. The force between them is given by the exact relation \cite{Griffiths07}
\begin{equation} \label{eq:force1}
	\vec{F}_{1,2}=
	\frac{\mu_0}{4\pi} \vec{\nabla} \cdot \left[  \frac{\vec{m}_1 \cdot \vec{m}_2}{r^3} - 3\frac{\left(\vec{m}_1 \cdot \vec{r} \right)\left( \vec{m}_2 \cdot \vec{r}\right)} {r^5}\right],
\end{equation}
\noindent where $\vec{r}$ is the vector joining the centres of the dipoles. This relation implies that, in general, the force depends on the angle between the dipoles. In the particular case when the dipole moments are equal in magnitude, parallel to each other, and perpendicular to $\vec{r}$ (dipoles in the same plane), the force takes the simpler form
\begin{equation} \label{eq:force2}
	\vec{F}_{1,2}=\frac{3\mu_0}{4\pi}\frac{m^2}{r^4} \hat{x},
\end{equation}
where $m=|\vec{m}_1|=|\vec{m}_2|$, $\mu_0$ is the permeability of the medium and $\hat{x}$ is an unitary vector in the direction of the axis that connect the centres of the magnets. Eq.~(\ref{eq:force2}) gives the force at equilibrium position ($r=a$) at a magnetic dipole ($n=1$) of the chain produced by is neighbour ($n=2$) in the chain. A opposite force is produced on the oscillator $n=2$.

In the case of the perturbed chain of magnets with nearest neighbour interactions, the distance between centres is a dynamic variable. Assuming small displacements of the magnets, i.e., the angles between the dipole moments are small, we can use Eq.~(\ref{eq:force2}) with $r = a-u_n+u_{n+1}$ to describe the interaction between two neighbour oscillators
\begin{equation} \label{eq:force3}
\vec{F}_{n,n+1}=\frac{3\mu_0 m^2}{4\pi}  \frac{1}{\left(a-u_n+u_{n+1}\right)^4}.
\end{equation}
\noindent Comparing with the equation of motion of the chain, given by Eq.~(\ref{eq:eqmotion2}), we identify the parameters
\begin{equation}
	\beta=(3/4\pi)\mu_0 m^2,\quad \alpha=4.
\end{equation}
This small angle Eq.~(\ref{eq:force3}) for the forces is a crude approximation and exact expressions can be found in Ref.~\cite{russell1997moving}. However, since our aim is to obtain simple analytical expressions based on the FPU equation, Eq.~(\ref{eq:FPU}), we will keep this degree of accuracy. The validity of this approximation to describe our setup will be tested in the next sections comparing with the experimental results and numerical simulations.

The above expressions for the forces between magnetic dipoles are valid for loop currents or magnets of negligible dimensions. Expressions for finite size magnets can be found in the literature \cite{Camacho13} and are in general lengthy and cumbersome. Gilbert's model of magnetic field of magnets used here results in approximate but simple expressions for the forces \cite{Griffiths07}. For cylindrical magnets of length $h$, with their magnetic moments parallel, and their axis perpendicular to the line joining the centres, the force between adjacent magnets can be expressed as
\begin{equation} \label{eq:force4}
\vec{F}_{1,2}=\frac{\mu_0 m^2}{2\pi h^2} \left( \frac{1}{r^2}- \frac{r}{\left(r^2+h^2\right)^{3/2}} \right) \hat{x},
\end{equation}
\noindent where the magnetic moment is $m={\cal{M}} h \pi R^2$, $\cal{M}$ is the magnetization and $R$ the radius of a the cylindrical magnet. In the limit $h\ll r$, Eq.~(\ref{eq:force4}) reduces to Eq.~(\ref{eq:force2}), i.e., magnets with small dimensions compared to their separation interact via dipolar forces, i.e., $\alpha=4$. In the opposite limit $h\gg r$, parallel magnets close to each other, the interaction law approaches to a Coulomb-type force, i.e., $\alpha=2$. In general, the interaction law of magnets can be approximated by an inverse-law with any given exponent that ranges between monopole and dipole cases.

\subsection{Experimental setup}
A chain of coupled magnets was built in order to test the theoretical predictions. The experimental setup is shown in Fig.~\ref{setup}. The chain was composed by 53 identical cylindrical neodymium magnets (Webcraft GmbH, DE, magnet type N45), with mass $M=2$ g, arranged in a one-dimensional periodic lattice. The radius and height of the magnets were $R=2.5$ mm and $h=14$ mm, respectively, and its magnetization was ${\cal{M}} = 1.07~10^6$ A/m. The magnets were oriented with the closest poles being those of the same polarity, therefore the produced forces were repulsive.

To achieve the necessary stability of the chain, the magnets were attached to a rigid bar which allows them to oscillate around a T-shaped support, being each magnet actually a pendulum (see Fig.~\ref{setup}). The length of the vertical bars was $L=100$ mm, and the distance between supports (and therefore the distance between magnets at equilibrium) was $a=20$ mm. The bearing of the T-shaped support was specially designed to minimize the effects of friction and giving stability to the system. This was achieved by using an additional ring-shaped magnets which keep the oscillators quasi-levitating on air, with just one contact point, as shown in the inset of Fig.~(\ref{setup}).

The effect of the pendulums is to introduce an additional external force to the dynamics of the chain, corresponding to the term $V'_\mathrm{ext}$ in Eq.~(\ref{eq:eqmotion2}). If $\theta_n$ is the angle formed by a magnet with respect to its vertical equilibrium position, the restoring force due to gravity is $F_z=M g \sin\theta_n$. For small angles $\theta_n$, and using the notation of Eq.~(\ref{eq:eqmotion2}), the force per mass can be approximated as $V'_\mathrm{ext}\simeq \Omega_0^2 u_n$, with $\Omega_0=\sqrt{g/L}/\omega_m$. 

All magnets oscillate freely except the outermost boundary magnets. The last magnet is fixed, and the first one is attached to the excitation system. The driving system consists of an electrodynamic sub-woofer (Fostex-L363) connected to an audio amplifier (Europower EPS2500) and excited by an arbitrary function generator (Tektronix AFG-2021). The first magnet is attached to the loudspeaker's diaphragm, thus, being it forced with a sinusoidal motion for different values of frequencies and amplitudes.

The motion of the chain is recorded by using a GoPro-Hero3 camera. The camera is placed at a proper distance from the chain in order to track the motion of a certain number of magnets. In this work, the first $18$ magnets were recorded simultaneously. Then, each pendulum was optically tracked using image post-processing techniques. Image calibration was employed here to correct the lens aberration using the image processing toolbox in Matlab\textsuperscript{\textregistered}, allowing the measurement of the displacement waveforms $u_n$. We considered the travelling wave regime, ignoring the reflected wave by time windowing the recorded video. The measurement in a finite time window guarantees no reflections from the $n=N$ boundary. Due to the quasi-instantaneous temporal duration of the impulse response of the system, after some temporal cycles of measurement the system reaches the stationary. Therefore, the transient measurement is equivalent to the response of an infinite chain and the finite size effects of the chain do not influence the experiments. 

The duration of each record was about $3.5$ s, the camera resolution was set to $960$p with a frame rate of 100 frames per second, i.e., leading to a sampling frequency of 100 Hz. Using the measured waveforms, the amplitude of each harmonic was estimated as usual using the Fourier transform.  

\section{Experimental results}\label{sec:results}
\subsection{Dispersion relation}

\begin{SCfigure*}[1][t]
	\centering
	\includegraphics[width=13cm]{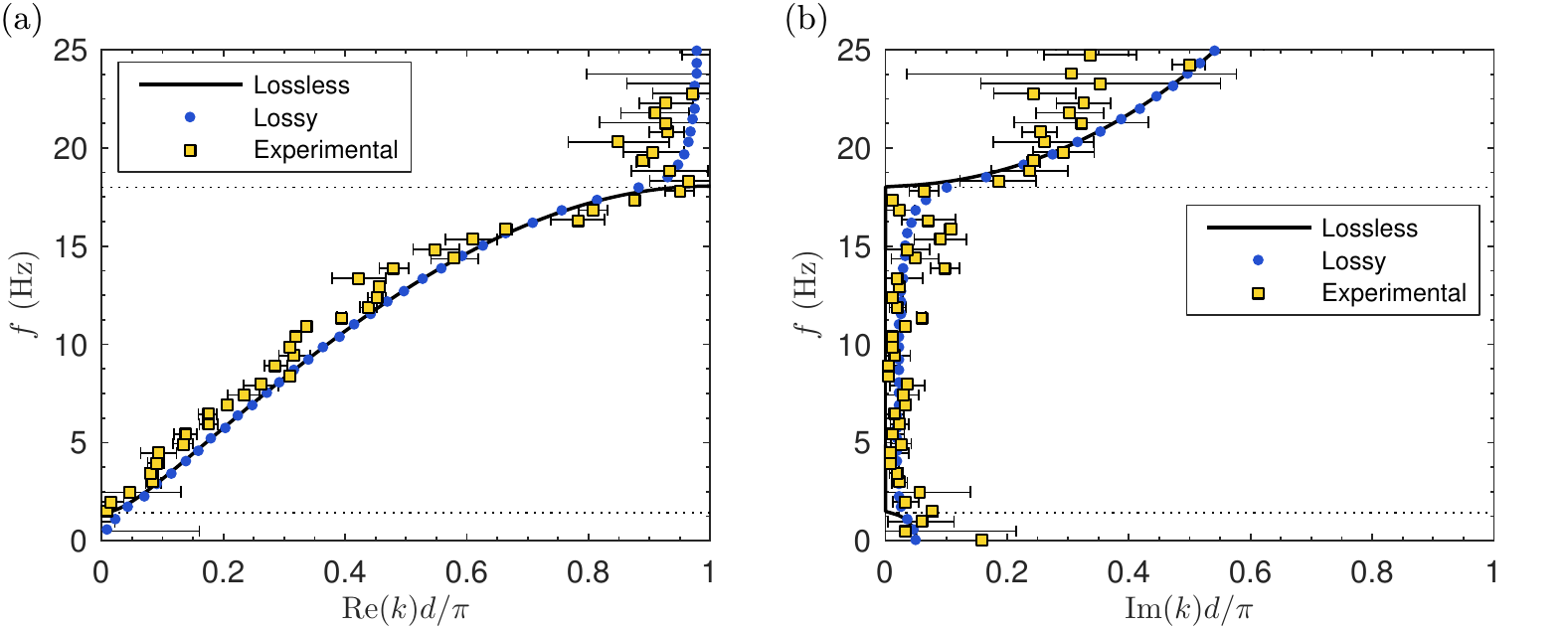}
	\caption{Dispersion relations of a mono-atomic chain obtained analytically by using Eq.~(\ref{disp2}) (continuous line), by the experimental measurements (squares), and numerically including damping (circles). Horizontal bars indicate the experimental error of the normalized wavenumber. (a) Real part of the wavenumber, (b) imaginary part of the wavenumber.}
	\label{Disp_Mono}
\end{SCfigure*}

To obtain the dispersion relation experimentally, the first magnet was excited with short duration impulse with low amplitude excitation in order to ensure that the excited waves are described by linear theory. The generated travelling pulse was recorded at two consecutive magnets, i.e., $n$ and $n+1$. The real part of the wavenumber was calculated by estimating the phase difference between them and the imaginary part of the wavenumber was calculated estimating the attenuation, as 
\begin{align}
 	\mathrm{Re}(k) &= \frac{\omega}{\mathrm{Re}(c_p)} = \frac{\arg [U_{n+1}(\omega)/U_n(\omega)]}{a}, \\
 	\mathrm{Im}(k) &= \frac{\omega}{\mathrm{Im}(c_p)} = \frac{\log\left|U_{n+1}(\omega)/U_n(\omega)\right|}{a}, 	
\end{align}
\noindent where $U_n(\omega)$ is the Fourier transform of the measured displacement of the $n$-th magnet and $c_p$ is the phase velocity. A set of 10 measurements at the oscillator $n=3$ were used to compute the mean value of the phase speed. Figures~\ref{Disp_Mono}~(a-b) show the real and imaginary part of the wavenumber respectively, where the experimental results and the dispersion relation of Eq.~(\ref{disp2}) were evaluated at frequencies with step of $\Delta f= 0.66$ Hz. The small magnitude of the experimental errors in the propagating band indicates good repetitiveness of the measurements. The experimental lower frequency cut-off was $f_0=1.68$ Hz, which agrees with the theoretical value $f_0=(1/2\pi) \sqrt{g/L} = 1.48$ Hz, ($f_0=1.56$ Hz if we consider the rigid-body pendulum taking into account the momentum of inertia of the steel rod). The measured upper cut-off frequency was $f_m=17.7$ Hz. This value was used to fit Eq. (\ref{eq:force4}) to an inverse power law, using the theoretical prediction $f_m=(1/2\pi)\sqrt{4 \alpha \beta/M a^{\alpha+1}} = 17.6$ we obtained an inverse power law with exponent $\alpha=3.6$ (quasi-dipolar interaction), which is in agreement with the ratio between the height and the separation distance between of the magnets given by Eq.~(\ref{eq:force4}). Both, upper and lower values of the dispersion relation obtained experimentally can slightly change with the amplitude of the input excitation $u_0$, which is in fact a signature of nonlinear dispersion caused by the finite amplitude of the wave. Note that for higher amplitudes the pulsed excitation used in this experiments leads to the generation of KdV-like compression solitons \cite{moleron2014solitary}. However, as long as the condition $u_0\ll a$ is fulfilled, the chain propagates linear modes and the dispersion relation can be obtained. 

One remarkable result is the low damping of the system, given by the smallness of the imaginary part of the wavenumber in the propagating band. 

The complex dispersion relation obtained by numerical integration of Eq.~(\ref{eq:eqmotion2})  adding a damping term $\gamma \partial u_n / \partial t$ to the equation of motion is shown in Figs~\ref{Disp_Mono}~(a-b). The damping coefficient, $\gamma$, was fitted to the experiments and corresponds to 0.52 dB/m (note the chain is 1 m long). The damping terms produces a force that opposes the pendulum movement. It is worth noting here that, in the propagating band, the total drag force is roughly twice the viscous drag force estimated for a cylinder of the size of a single magnet oscillating in air \cite{brouwers1985}: the magnetic bearing system itself produces only small damping. The effect of the small losses is to smooth the limits of the band gap, as it is also observed in other highly dispersive systems, e.g., as in Acoustics \cite{jimenez2017}, and to produce a small attenuation in the propagating band. The damping term is used in the numerical simulations in the following sections. 

\begin{SCfigure*}[1][tp]
	\centering
	\includegraphics[width=13cm]{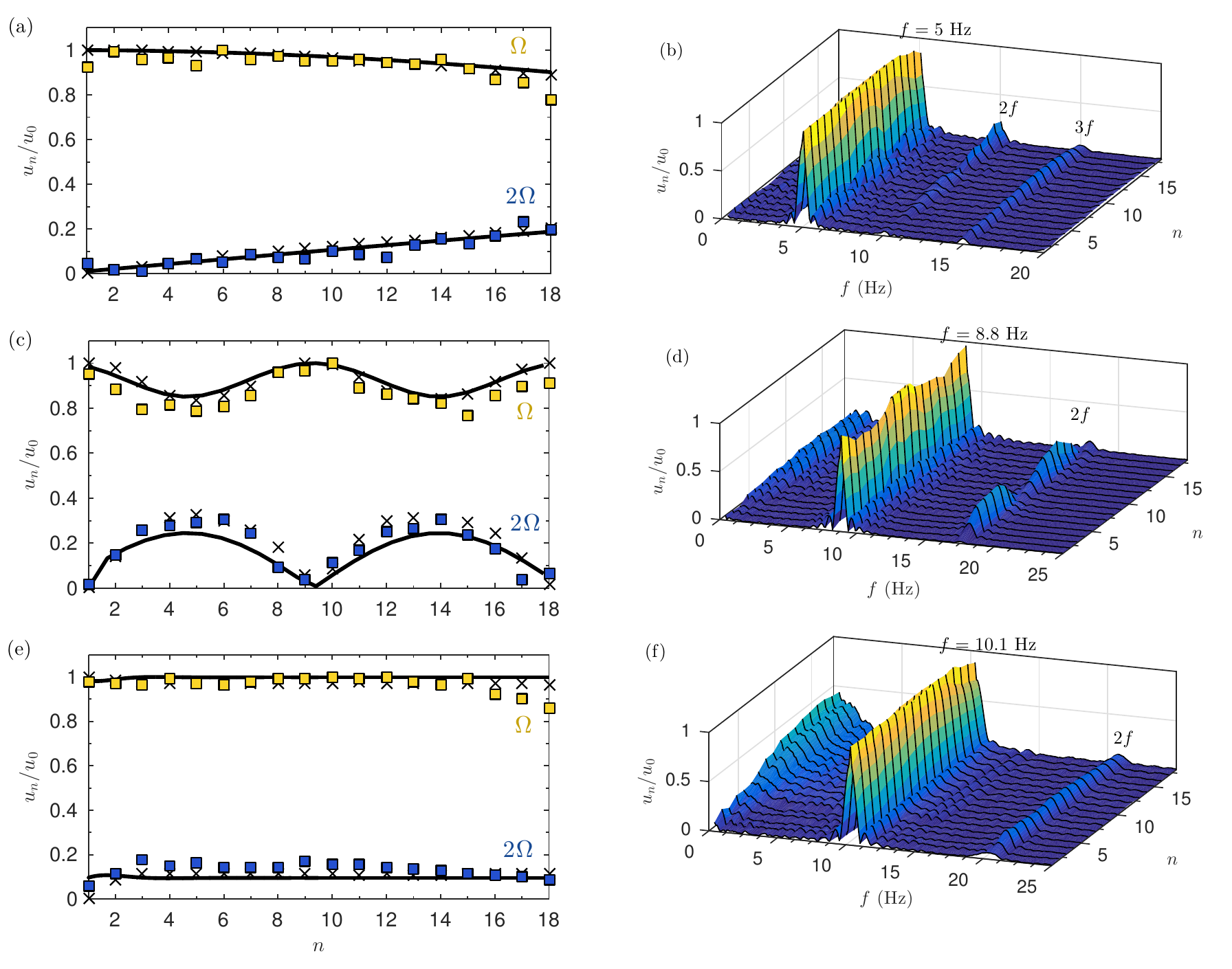}
	\caption{Three different regimes of harmonic generation, measured at different frequencies. (a) Weakly dispersive regime ($f=5$ Hz, $\Omega=0.27$) obtained using the analytical equation (continuous lines), numerical solution of the motion equations (crosses) and experimental results (squares). (b) Corresponding experimental spectrum as a function of the oscillator number. (c) Strongly dispersive regime ($f=8.8$ Hz, $\Omega=0.48$), and (d) its corresponding spectrum. (e) Evanescent regime for the second harmonic ($f=10.1$ Hz, $\Omega=0.55$), (f) corresponding spectrum.}
	\label{fig-harmonics} 
\end{SCfigure*}

\subsection{Harmonic generation }

By driving the first magnet with a sinusoidal motion, $u_1=u_0 \sin \omega t$, harmonic waves are excited and they propagate along the chain. On one hand, the amplitude $u_0$ was $u_0 = 2.4$ mm. On the other hand, according to the dispersion relation shown in Fig.~\ref{Disp_Mono}, the driving frequency, $\Omega$, can be chosen among to three different regimes regarding the propagation of the second harmonic: (a) weakly dispersive, (b) strongly dispersive, and (c) evanescent.

The first case (a) is obtained when the frequency of the fundamental wave lies in the lower part of the pass band and, the generated second harmonic is also in the pass band, in the region of weak dispersion. Thus, in this regime the motion equations of the lattice can be approximated by a continuum whose dynamics follows the Boussinesq equation \cite{sanchez2013second} and the wave roughly propagates without dispersion. In this low frequency regime, the lower harmonics propagate with nearly the same phase velocity. The amplitude of the second harmonic increases roughly linearly with distance while the first harmonic amplitude decreases due to the energy transfer from the fundamental component to the higher harmonics. This case is shown in Fig.~\ref{fig-harmonics}~(a), where a fundamental wave with frequency $f=5$ Hz ($\Omega=0.27$) generates a second harmonic whose frequency $2f=10$ Hz lies on the weakly-dispersive region of the propagative band ($\Omega=0.54$). We note that in this regime, third harmonic is also generated, as shown Fig.~\ref{fig-harmonics}~(b), although it is not predicted by the perturbative analytical solution due to its second-order accuracy.

Secondly, the case (b), corresponding to strongly dispersive second harmonic, is shown in Fig.~\ref{fig-harmonics}~(b). Here, the driving frequency approaches the half of the passband frequency. The second harmonic lies in the highly dispersive part of the band, but still in a propagative region (slightly below the cutoff frequency $f_m$). As observed in the previous case, the amplitude of the second harmonic increases with distance, but now at a particular distance given by the coherence length, $l_c$, it decreases. Both, the fundamental wave and its second harmonic present spatial oscillations, i.e., spatial beatings. Figures~\ref{fig-harmonics}~(c-d) illustrate this case for a fundamental wave with frequency $f=8.8$ Hz, i.e. a second harmonic with frequency $\Omega=0.96$. The experimental value  of the coherence length was $l_c \approx 4.5a$, which is in agreement with the theoretical given by Eq.~(\ref{coherence}).

Finally, the case (c) corresponds to the second harmonic lying within the band-gap, as shown in Fig. \ref{fig-harmonics}~(e-f) for a excitation frequency of $f=10.1$ Hz ($\Omega=0.55$). In this case, the second harmonic is evanescent and its amplitude does not change with distance. One would expect the absence of the SH field (since it is an evanescent mode) but a finite amplitude is observed in agreement with theory and numerical simulations. The second harmonic component is generated locally as it is ``pumped" by the fundamental wave. Its amplitude value remains constant all along the chain, being its amplitude dependent on the driving amplitude and on the properties of the medium (the non-linearity and the magnitude of the dispersion).  

The experimental results shown in Fig.~\ref{fig-harmonics} are in good agreement with the analytical predictions of the asymptotic theory (solid lines), and also with the numerical simulation of Eq. (\ref{eq:FPU}). However, small discrepancies can be observed between the theory and the experiments, as well as between the theory and the simulations. The value of the nonlinear coefficient used in the experiments was $\varepsilon=(1+\alpha)u_0 = 0.55$. Thus the small disagreements between the theory and the experiments and simulations are mainly explained due to the non smallness of the nonlinear parameter $\varepsilon$. For small excitation amplitudes, i.e., small $\varepsilon$, the theory and numerical solutions converge to the similar result. However, due to the precision of the motion-tracking acquisition system, it was difficult to accurately measure small amplitude perturbations.
 \begin{SCfigure*}[1][t]
 	\centering
 	\includegraphics[width=14cm]{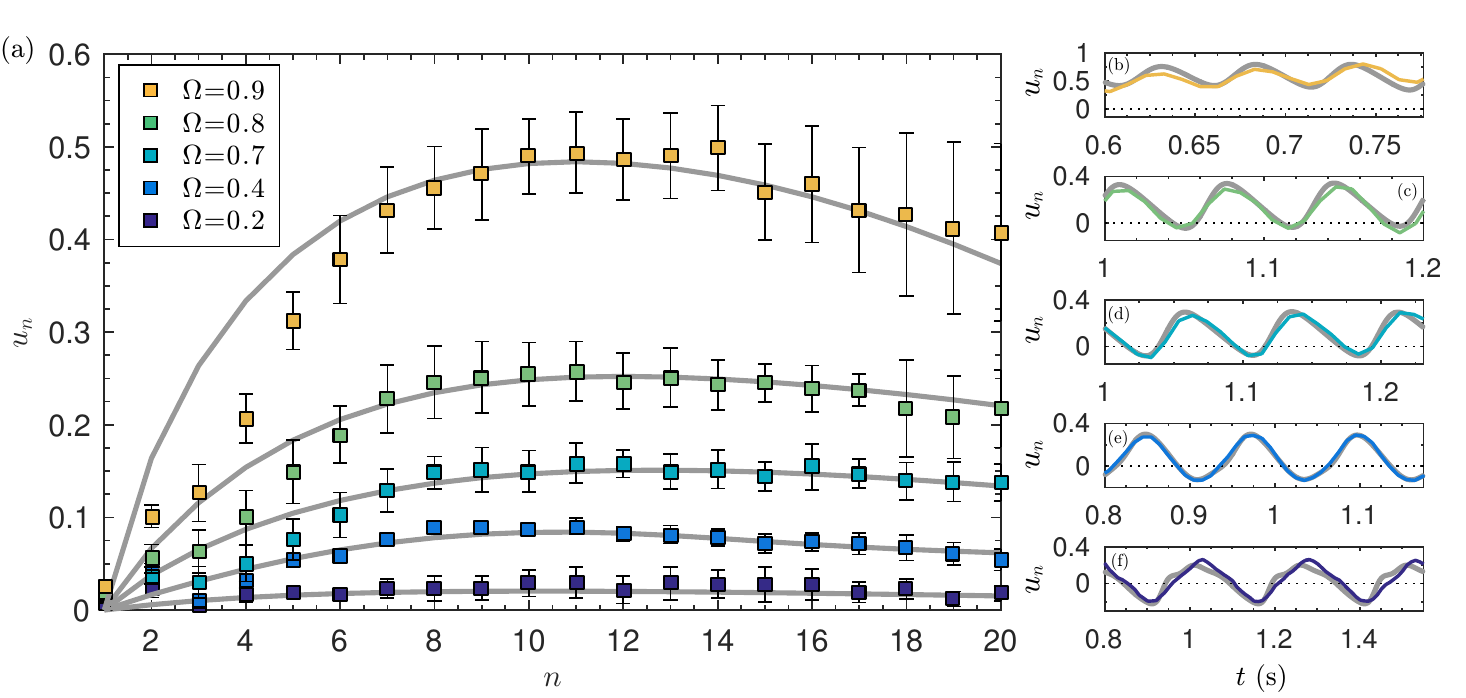}
 	\caption{(a) Amplitude of the static displacement mode as a function of the space obtained by numerical integration of the equation of motion (continuous lines) and measured experimentally (markers) at different frequencies. (b-e) Corresponding experimental (coloured lines) and simulated (grey) waveforms acquired at oscillator $n=14$.}
 	\label{expansion}
 \end{SCfigure*}

\subsection{Chain dilatation}
Besides the harmonic generation, the FPU equation also predicts the presence of a static (zero-frequency) mode. It physically represents an incremental shift of the average position of each oscillator, which in turn results in a constant dilatation or expansion of the chain. This term is accounted for by the first term in Eq.~(\ref{eq:Analytic}). Since the average displacement grows linearly with distance, it can be interpreted as a constant strain produced by the acoustic mode along the lattice.

The phenomenon was originally reported for acoustic waves propagating in a solid described by a nonlinear wave equation \cite{cantrell1991acoustic}, which is actually the continuous (long-wavelength) analogue of Eq.~(\ref{eq:FPU}). The effect was described there as an acoustic-radiation-induced strain. The physical origin of the expansion of the discrete chain (and also in the continuous solid) is the anharmonicity of the interaction potential, and therefore is a general nonlinear effect. Note radiation forces also appear in other nonlinear systems as acoustic waves in fluids, soft solids or even light (radiation pressure), being the generation of acoustic radiation forces a general mechanism of any wave motion \cite{sarvazyan2010}. We remark that the phenomenon of acoustic expansion is analogous to the thermal expansion of solids, which also has its physical origin in the lattice anharmonicity. The link between these two effects and its relation with the acoustic nonlinear parameter has been pointed out in Ref.~\cite{Cantrell82}.  

We have shown in Fig.~\ref{expansion} the generation of the zero-mode in the particular case of the chain of coupled oscillators and for different excitation frequencies. The experimental results agree with simulations of the full equations of motion including the restoring force. We can see that for all the frequencies the linear increasing of the displacement predicted by the analytical solutions is not observed. Instead, we can observe two regimes, and a transition between them at a particular distance. First, in the region near the boundary (extending up to $n\approx 8$ in our experiment), the displacement grows roughly linearly with distance, as predicted by the theory without restoring force. However, beyond a given distance the growth of the static displacement mode saturates, and the chain attains an unstrained state, with the oscillators moving around positions shifted with respect to their initial values. This behaviour is not predicted by the theory. 

The saturation effect can be understood if we recall that the theory was developed assuming that there was not a prescribed equilibrium position for any oscillator in the chain, the chain was assumed semi-infinite, and the only force acting on the masses was the nearest neighbours interaction. However, in the experimental setup an additional restoring force is present, due to gravity. For small perturbations this is equivalent to an on-site potential. Since the magnets are pendula, the maximum shift of a magnet with respect to the equilibrium position is also bounded. Note in Fig.~\ref{expansion} the oscillators are displaced less than a lattice step. Note also that for a finite value of the on-site potential $\Omega_0$ the zero-th mode is always evanescent. Then, as described previously with the second harmonic in the evanescent case, only the forced contribution to the zero-th order mode is present, leading to a constant value of the zero-th mode. 

\begin{figure*}[t]
 	\centering
 	\includegraphics[width=1\textwidth]{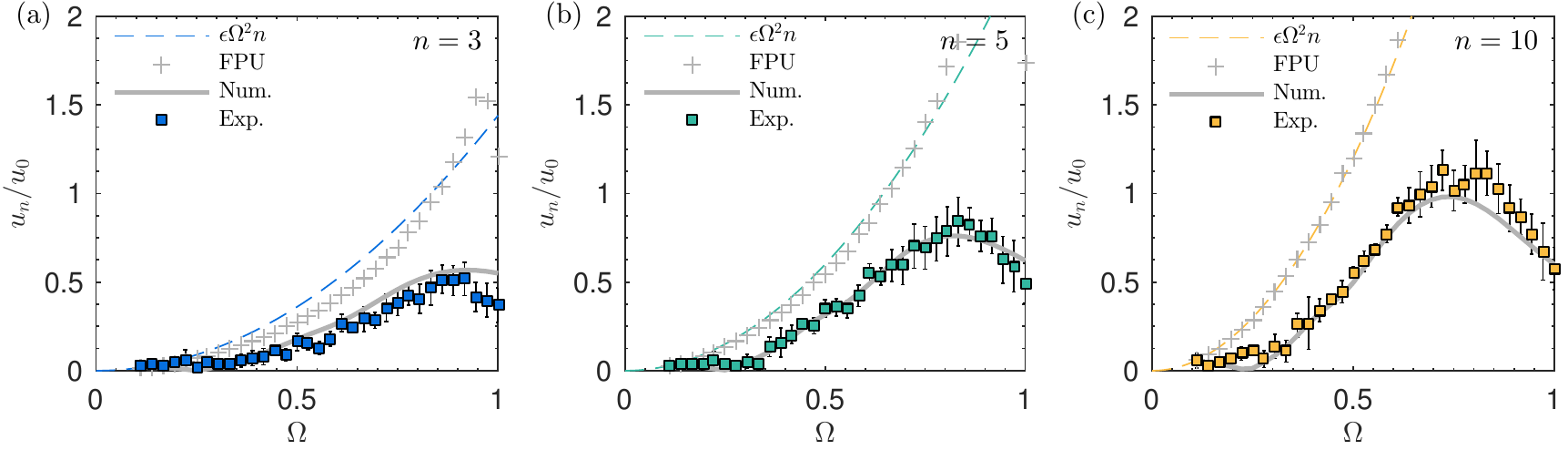}
 	\caption{Dependence of the amplitude of static mode with the excitation frequency obtained using the analytical solution (dashed lines), numerical integration of the FPU equation (crosses), numerical integration with the pendulum restoring force (grey thick line) and experiments (squares), measured at the oscillator (a) $n=3$, (b) $n=5$ and (c) $n=10$.}
 	\label{expansion2}
 \end{figure*}

Finally, Fig.~\ref{expansion2} shows the dependence of the zero-mode with frequency, measured at $n=3$, $n=5$ and $n=10$. It can be observed that the experimental results agree with the simulations of the full equations of motion, while the simulations of the FPU equation roughly does with the theory (in this case the excitation amplitude was $u_0=4.8$ mm, leading to a value of the nonlinear parameter of $\varepsilon=0.96$). For frequencies below $\Omega\approx0.8$, the amplitude of the zero-mode roughly follows a quadratic dependence with frequency. In addition, the period average displacement of an oscillator corresponds to the position where there exist a balance between the gravity restoring force and the equivalent acoustic-radiation force produced by the nonlinear compressional wave, corresponding to $F_\mathrm{ARF} = \Omega_0^2 \left\langle u_n\right\rangle$, where $\left\langle u_n\right\rangle$ is the amplitude of the zero mode. Thus, the induced acoustic-radiation force in the experimental chain also follows a quadratic dependence with frequency for low frequency waves.
\section{Conclusions}\label{sec:conclusions}
The propagation of nonlinear monochromatic waves in a lattice of particles coupled by repulsive forces following an inverse power-law with distance has been studied theoretically, numerically and experimentally. In the limit of small amplitudes, the system is described by a FPU equation with quadratic nonlinearity, where analytical solutions were generalized for the case of an arbitrary inverse frequency power-law interactions. In particular, it has been developed an experiment consisting in a lattice of coupled magnetic dipoles sinusoidally driven at one boundary, while a magnetic bearing system for the rotation of each pendulum provides low mechanical damping. 

In spite of the simplifying assumptions made in the theoretical analysis, the observations agree quite well with the model concerning the generation of the second harmonic, e.g., the characteristic spatial beatings of the second harmonic due to the dispersion of the lattice are observed. 

One particular feature of the studied lattice is the existence of a restoring force due to the action of gravity on the pendula. This is roughly equivalent to the introduction of an on-site potential, leading to the generation of a low frequency band gap. In this work, it has been observed for the first time that the generated zero mode is evanescent due to the presence of the on-site potential, therefore, only the forced component of the zero mode propagates through the chain and a saturation of the amplitude of the zero mode is observed. There exist discrepancies between the analytical FPU theory and the experimental measurements of the static dilatation mode. They are caused, mainly, because the developed theory is based on a FPU equation that lacks of the on-site potential that produces the low frequency band gap. Therefore, while the FPU theory predicts a linear monotonic growth of the zero-mode, the presence of the low-frequency band gap makes the zero-frequency mode to be evanescent, and, as a consequence, a saturation of the dilatation of the chain is observed in the experiments and in the numerical simulations. The particular dynamics of the generated zero-mode are discussed in analogy with the radiation force produced by a nonlinear monochromatic travelling wave. This result has an interest beyond the particular studied system, since there exist a number of systems, e.g. as condensed matter or granular crystals, that present similar of dispersion relations, with a low-frequency band gap.

Additionally, the present low-friction experimental setup can be used to explore other effects of nonlinear discrete systems that have been predicted in the literature, e.g., nonlinear localized modes. Under the assumption of small amplitude, these results indicate that the lattice of magnetic dipoles is well described by an $\alpha$-FPU equation, which opens the possibility of extending the results to other systems which are described by the same generic equation. The proposed system can be also viewed as a mechanical analogue of a microscopic crystal of interacting charged particles (atoms or ions) at a macroscopic scale. Despite the limited applicability of this simple one dimensional lattice to describe real crystals, the approach possess however some advantages, as the possibility of varying parameters that are normally fixed, as the strength of the interaction and on-site potentials, or exploring strongly nonlinear regimes which are hardly achievable at atomic scales.

\acknowledgments
This research was funded by Spanish Ministerio de Economia e Innovacion (MINECO), grant  FIS2015-65998-C2-2-P. AM gratefully acknowledge to Generalitat Valenciana (Santiago Grisolia program). LJSC gratefully acknowledge the support of PAID-01-14 at Universitat Polit\`ecnica de Val\`encia.


\end{document}